\documentclass[12pt]{article}
\usepackage{graphicx}
\usepackage{bm}        % for math
\usepackage{amssymb}   % for math

\def\beq{\begin{eqnarray}}
\def\eeq{\end{eqnarray}}

\def\ba{\begin{eqnarray}}
\def\ea{\end{eqnarray}}
\def\beq{\begin{eqnarray}}
\def\eeq{\end{eqnarray}}

\def\mpl{M_{\rm Pl}}

\def\d{\mathrm{d}}
\def\p{{\cal P}}

\def\L*{{\cal L}_*}
\def\L{\mathcal{L}}
\def\({\left(}
\def\){\right)}

\def\nn{\nonumber}
\def\p{\partial}
\def\mn{_{\mu \nu}}
\def\stu{St\"uckelberg }
\def\p{\partial}

\def\<{\langle}
\def\>{\rangle}
\def\K{{\cal K}}

\newcommand{\de}{\partial}
\newcommand{\be}{\begin{equation}}
\newcommand{\ee}{\end{equation}}
\newcommand{\al}{\alpha}
\newcommand{\bt}{\beta}

\newcommand{\del}{\delta}

\newcommand{\lam}{\lambda}

\newcommand{\sig}{\sigma}

\newcommand{\Om}{\Omega}

\newcommand{\rmd}{\mathrm{d}}

\newcommand{\Mpl}{M_{\textrm{Pl}}}

\renewcommand{\d}{\mathrm{d}}

\newcommand{\x}{\vec{x}}

\newcommand{\Tr}{\mathrm{Tr}}

\def\lsim{\mathrel{\rlap{\lower3pt\hbox{\hskip0pt$\sim$}}
     \raise1pt\hbox{$<$}}}         %less than or approx. symbol
\def\gsim{\mathrel{\rlap{\lower4pt\hbox{\hskip1pt$\sim$}}
     \raise1pt\hbox{$>$}}}         %greater than or approx. symbol

\def\lsim{\mathrel{\rlap{\lower3pt\hbox{\hskip0pt$\sim$}}
     \raise1pt\hbox{$<$}}}         %less than or approx. symbol
\def\gsim{\mathrel{\rlap{\lower4pt\hbox{\hskip1pt$\sim$}}
     \raise1pt\hbox{$>$}}}         %greater than or approx. symbol

\usepackage{amsmath}
\usepackage{amsfonts}
\usepackage{verbatim}
\begin{document}

\begin{titlepage}

\begin{flushright}
{NYU-TH-08/28/11}

\today
\end{flushright}
\vskip 0.9cm

\centerline{\Large \bf Massive   Cosmologies}
\vskip 0.7cm
\centerline{\large G.~D'Amico$^a$, C.~de Rham $^{b,c}$, S.~Dubovsky$^a$, 
G.~Gabadadze$^a$,} 
\vspace{0.1in}
\centerline{\large D.~Pirtskhalava$^a$, A.~J.~Tolley$^c$}

\vskip 0.3cm

\centerline{\em $^a$Center for Cosmology and Particle Physics,
Department of Physics,}
\centerline{\em New York University, New York,
NY, 10003, USA}

\centerline{\em $^b$D\'epartment de Physique  Th\'eorique and Center for Astroparticle Physics,  }
\centerline{\em Universit\'e de  Gen\`eve, 24 Quai E. Ansermet, CH-1211  Gen\`eve}

\centerline{\em $^c$Department of Physics, Case Western Reserve University, Euclid Ave,
Cleveland, OH, 44106, USA}

\centerline{}
\centerline{}

\vskip 1.cm

\begin{abstract}

We explore the cosmological solutions of a recently proposed extension of General
Relativity with a Lorentz-invariant mass term. We show that the same constraint 
that removes the Boulware-Deser
ghost in this theory also prohibits  the existence  of homogeneous and isotropic cosmological
solutions.  Nevertheless,  within domains of the size of inverse graviton mass 
we find approximately homogeneous and isotropic solutions that can well describe 
the past and present of the Universe.  At energy densities above a certain 
crossover value,  these solutions approximate the standard FRW evolution with great accuracy.
As the Universe evolves and density drops below the crossover value 
the inhomogeneities become more and more pronounced. In the 
low density regime each domain  of the size of the inverse graviton mass has 
essentially non-FRW cosmology. This scenario  imposes an upper bound on the graviton mass, which
we roughly estimate to be an order of magnitude below the present-day value of the
Hubble parameter. The bound  becomes especially restrictive if one utilizes
an exact self-accelerated solution that this theory offers.

Although the above  are robust predictions of massive
gravity with an explicit mass term, we point out that if the mass parameter
emerges from some additional scalar field condensation, the constraint no longer
forbids the  homogeneous and isotropic cosmologies. In the latter
case, there will exist an extra light scalar field at cosmological scales,
which is screened by the Vainshtein mechanism at shorter distances.

\end{abstract}

%\vspace{3cm}

\end{titlepage}

\newpage

\section{Introduction and Summary}

The purpose of this work is to study the cosmology of General Relativity (GR)
with an explicit Lorentz invariant mass term (massive gravity or massive GR). Historically, it
has been difficult to construct a nonlinear theory of massive gravity that would describe no
more than the five degrees of freedom,
required  for the massive spin-2 state by  the representations of the
Poincar\'e group. This situation was recently transformed by the proposal in Ref.
\cite {drgt} of a theory of massive gravity with 5 degrees of freedom.

This theory was shown to be free of the sixth degree of freedom (the so-called
Boulware-Deser (BD) ghost \cite{bd})  to all orders in the decoupling limit (DL) in Refs.
\cite {drgt,drg2}, where it was also shown to be  ghost-free away from the DL
up to and including quartic order in nonlinearities \cite {drgt}.
These arguments were recently generalized in \cite{hr} to a complete
nonlinear proof of the absence of ghosts, away from the decoupling limit, in the
ADM/Hamiltonian formalism. The same result can be reached  in both the \stu
(see Ref.~\cite{drgt_res2}) and helicity formalisms (see Ref.~\cite{helicity}).
In each language, there exists a constraint that eliminates one
degree of freedom which otherwise would have been the BD ghost.

In this work we will show that the very same constraint that eliminates the BD ghost
in  massive gravity \cite{drgt},
also forbids homogeneous and isotropic cosmological solutions (FRW cosmologies).
For solutions with FRW symmetries, the all-orders constraint can straightforwardly
be seen to prohibit any time evolution, leaving Minkowski space as the only vacuum solution
which is consistent with homogeneity and isotropy.

This result raises the question: How could the non-FRW cosmologies of massive
gravity recover the FRW solutions of GR in the massless limit?
For this one should rely upon the Vainshtein  mechanism by which
massive GR is expected to recover GR in the $m\to 0$ limit \cite {Arkady}.
Although the original theory in which the Vainshtein mechanism was proposed contains a BD ghost,
the mechanism itself seems to be universal, and has been established in other models where the BD
ghost is not present \cite {DDGV,lpr,nr,nrt}.  Moreover, it was shown that the mechanism
is operative for spherically symmetric solutions in the massive GR theories discussed here,
at least for a certain choice of the three {\it a~priori} free  parameters of the theory
(i.e., the graviton mass and two arbitrary constants), \cite {knt,cp}.

Assuming that the Vainshtein mechanism is at work, one would expect to find
in massive GR cosmological solutions that are more and more homogeneous and isotropic
as the value of the graviton mass is taken to zero. If this is the case, then the fact that
massive gravity leads to non-FRW solutions will not immediately rule it out via observations,
but rather just place a constraint on the magnitude of the mass of the graviton, to be consistent
with known constraints on homogeneity and isotropy.

To see a  close connection between the Vainshtein mechanism and cosmology,
consider matter of constant density $\rho$ stored in a sphere of radius $R$.
The Vainshtein radius  of such a source is
\beq
r_* = \left ( {r_g\over m^2} \right )^{1/3}= \left ( {\rho \over 3 \mpl^2  m^2 } \right )^{1/3} R\,,
\label{r*}
\eeq
where $r_g = 2 MG_N$ is the gravitational radius of the  source of mass $M= {4\over 3}\pi R^3\rho$,
and $G_N=(8 \pi \mpl^2)^{-1}$.   Furthermore, it is useful to introduce a notion of a
crossover energy density,
\beq
\rho_{\rm co} \equiv 3 \mpl^2 m^2\,.
\label{rhoco}
\eeq
From (\ref {r*}) we conclude  that any source with density above the crossover value (\ref {rhoco}),
is characterized by the Vainshtein radius that is  greater than the size of the source itself.
For such sources gravity is close to that described by GR at distance scales $\ll r_*$ (the Vainshtein regime),
however, deviates significantly from GR at distance scales $\gsim r_*$ (the vDVZ regime) \cite {vdvz}.
Furthermore,  sources with density below the crossover
$\rho_{\rm co}$, are always in the vDVZ regime and their gravity differs
significantly from GR.

Let us now apply these observations to cosmology.  Suppose we took a  snapshot of a
universe at a certain stage of its evolution  when  the matter  in it
had an average energy density $\rho$ (averaged, say, at scales greater than the
Hubble scale $H^{-1}$ at that epoch; here we suppose that inflation,  or an alternative early
universe framework, prepared such a state). Let us look into a  $1/m$-size domain. 
An arbitrary Hubble patch in this domain 
(i.e.,  a patch enclosed by a sphere of radius $H^{-1}= (\rho/3\mpl^2)^{-1/2}$), that  
is far enough from the edges of the domain, is well-within the Vainshtein regime  
as long as $\rho \gg \rho_{\rm co}$. Then, cosmology  within  such Hubble  
patches can be approximated by the standard  FRW metric of GR 
with small corrections. Hence, for $\rho \gg \rho_{\rm co}$,  and 
well-within each $1/m$-size domain,  the early universe in massive GR
would  evolve as it  does in GR, with some  small corrections that  
vanish in the $m\to 0$ limit. Such an  expansion in each of these   
Hubble patches will last until the size of the patch, $H^{-1}$, 
approaches the scale $\sim 1/m$, or equivalently, until  $\rho$ dilutes 
down to density of the order of $\rho_{\rm co}$.  At scales  larger than $1/m$ 
gravitational interactions are expected to be screened.

The same arguments should hold for the radiation dominated epoch,
in which case $\rho \sim T^4$ ($T$ being temperature) should 
be compared with $\rho_{\rm co}$.

Requiring that the graviton mass be less then
the Hubble parameter today, $m< H_0$,  we find that $\rho_{\rm co}< \rho_c$, 
where $\rho_c$, is the present-day value of the critical density in the Universe. 
If so, then according to the above described scenario, the cosmological evolution
of the early Universe ($\rho \gg \rho_{\rm co}$) within each $1/m$-size 
domain will mimic the  FRW expansion with some  accuracy.
However, this will change significantly at densities $\rho \sim  \rho_{\rm co}$.
As long as the graviton mass $m$ is sufficiently small, the observational tests
of such cosmologies of massive GR would impose an upper bound  on $m$.
We estimate this bound to be approximately an order of magnitude smaller
than $H_0 \simeq 10^{-33}~{\rm eV}$.

\vspace{0.1in}

In general, the mass term introduces an effective stress-tensor in the
Einstein equation.  The backreaction of this term should be negligible
in the Vainshtein regime,  becoming dominant in the
vDVZ regime.  While generically we expect the above to be the case,
interestingly enough, we find one particular exact solution for which the backreaction
is described by a perfect fluid with the equation of state of dark energy,
and the magnitude of the energy density/pressure set by $m^2\mpl^2$. This behavior
is similar to the self-accelerated solutions of massive GR first found in Ref. \cite{drghp}
in the DL or its extension in Ref.~\cite{covariant}, and to exact self-accelerated solutions 
obtained in Refs. \cite{knt,theo,volkov}.

If one utilizes this particular solution, then by the end of the Vainshtein
regime the universe may become dominated by the self-accelerated solution.
However, the latter is not regular at spatial infinity, and could only exist
as a transient solution in space and time, matched upon the low-density
inhomogeneous  solution at larger  scales\footnote{Note that the solutions 
of Refs. \cite{knt,theo} exhibit singularities at  finite 
values of the coordinates, and hence,  should be matched to other 
solutions before reaching those points.}. 
Whether such a matching is possible, is not shown here; in principle 
the  evolution could  just bypass this solution and transition directly 
to a low density regime. Putting the question about the matching  aside, however,  
the bound on the graviton mass becomes especially restrictive if the expansion of 
the universe is described by the self-accelerated solution (see discussions in section 3).

In this paper we consider   the
cosmological evolution both in the Vainshtein and vDVZ regimes.
In the Vainshtein regime,  the metric to which the matter couples is
homogeneous and isotropic with some small corrections, but  it's the \stu sector that carries all 
the  inhomogeneities. Moreover, in this regime,
and for the self-accelerated solution, we will show the existence of the backreaction of the
mass term that is small and mimics dark energy. What is not shown is that there is a matching between
the Vainshtein (with or without self-acceleration) and vDVZ  regimes\footnote{To address this issue,
one could consider a possibility that the matter/radiation that is being  expelled from the bulk of 
the $1/m$-size domains, which are densely packed and adjacent to each other,  
gets  accumulated  near the boundaries of the domains. If the domains are 
well separated, or there is only one domain, density near the edge 
will be suppressed due to screening of gravity and free streaming. 
Different scenaria are determined by different initial conditions and need more detailed studies.}.

\vspace{0.1in}

The theory of Ref.  \cite {drgt} is the only potentially viable
classical  theory of Lorentz invariant massive GR with 5 helicity
states. Its cosmology is unusual, and this paper is a first attempt
at unfolding peculiarities of such a theory in a cosmological setup.  Therefore,  the majority
of this paper is qualitative in character, where we emphasize certain
universal aspects  and set up a general framework in which such cosmologies can
further be studied in details. A number of particular exact cosmological solutions
are discussed in the appendices.

\section{Massive GR and Cosmology}

For massive GR  the action is a functional of the
metric $g_{\mu\nu}(x)$,  and four spurious scalar fields
$\phi^a(x),~a=0,1,2,3$;  the latter are introduced to give a manifestly 
diffeomorphism invariant description \cite{ags,sergei}.
One  defines a covariant tensor $H_{\mu\nu}$ as follows:
\beq
g_{\mu\nu} = \partial_\mu \phi^a
\partial_\nu \phi^b \eta_{ab} + H_{\mu\nu}\,,
\label{gH}
\eeq
where $\eta_{ab}={\rm diag}(-1,1,1,1)$. The first term on the r.h.s. is
nothing but the Minkowski metric in the coordinate system defined
by $\phi^a$'s. Hence, gravity in this formulation is described by
the tensor  $H_{\mu\nu}$ propagating on Minkowski space.
In the unitary gauge all the
four  scalars $\phi^a(x)$ are frozen and equal to the corresponding space-time
coordinates,  $\phi^a(x)= x^\mu \delta_\mu^a$. However,
often it is helpful to use a non-unitary gauge in which $\phi^a(x)$'s
are allowed to fluctuate.

A covariant Lagrangian density for massive GR
can be written as follows,
\beq
\mathcal{L}=\frac{\mpl^2}{2} \sqrt{-g} \(R-\frac{m^2}{4}\,
\mathcal{U}(g,H)\)\,,
\label{L2}
\eeq
where $\mathcal{U}$ includes the mass,  and non-derivative
interaction  terms for $H_{\mu\nu}$ and  $g_{\mu\nu}$ .

A necessary condition for the theory to be ghost free in the DL
is that  the potential  $\sqrt{-g}\ \mathcal{U}(g,H)$  be
a total derivative upon the field substitution $h_{\mu\nu}\equiv g_{\mu\nu} -\eta_{\mu\nu}=0,~~
\phi^a = \delta_\mu^a x^\mu - \eta^{a\mu}\partial_\mu\pi$  \cite {drg2}.
With this substitution, the potential $\sqrt{-g}\ \mathcal{U}(g,H)$  becomes a function of
$\Pi_{\mu\nu} \equiv \partial_\mu  \partial_\nu \pi $ and its various contractions.

For instance, the following expression composed of  $\Pi_{\mu\nu}$
is a total derivative
\beq
\label{L2der}
\mathcal{L}^{(2)}_{\rm der}(\Pi) &\equiv &[\Pi]^2-[\Pi^2]\,,
\eeq
where we use the notations  $[\Pi] \equiv \operatorname{tr} \Pi^\mu_\nu $,
$[\Pi]^2 \equiv (\operatorname{tr} \Pi^\mu_\nu )^2$, while
$ [\Pi^2 ] \equiv \operatorname{tr} \Pi^\mu_\nu \Pi^\nu_\alpha$.

Then, as argued in \cite {drgt}, the Lagrangian for massive GR
that is automatically  ghost free  to all
orders in the DL  is obtained  by replacing the matrix elements
$\Pi^\mu_\nu $ in the total derivative term  (\ref {L2der})
by the matrix elements of a tensor $\K^\mu_\nu$,  defined as follows:
\beq
\label{Kmn}
\K^\mu_\nu (g,H)\,=\,
\delta^\mu_\nu -\sqrt{\partial^\mu \phi^a \partial_\nu \phi^b \eta_{ab}}\,.
\eeq
Here, the indices on $\K$ should be
lowered and raised by $g_{\mu\nu}$ and its inverse respectively.
This procedure defines the mass term (along with the interaction
potential in the Lagrangian density)  in
massive GR
\beq
\mathcal{L}= \frac{\mpl^2}{2} \sqrt{-g} \left[ R- m^2 \(  \K^\mu_\nu \K^\nu_\mu -
(\K^\alpha_\alpha )^2 \) \right] .
\label{L2K}
\eeq
The matter and other fields are coupled to $g_{\mu\nu}$ as in GR.
The above expression has no free parameters once  the graviton mass
is fixed.  In general, however, there  exist other polynomial
terms  in $\K$ with similar properties. These terms can be constructed
straightforwardly by using the procedure outlined in Ref. \cite {drgt}.
In  any dimensions there are only a finite number of
total derivative combinations, made of $\Pi$ \cite {nrt}.  They are all
captured  by the recurrence relation \cite {drg2}:
\beq
\label{Ldern}
\mathcal{L}_{\rm der}^{(n)}=-\sum_{m=1}^{n}(-1)^m\frac{(n-1)!}{(n-m)!}\,
[\Pi^{m}]\,\mathcal{L}^{(n-m)}_{\rm der}\,,
\eeq
with $\mathcal{L}^{(0)}_{\rm der}=1$ and $\mathcal{L}^{(1)}_{\rm der}=[\Pi]$.
This also guarantees that the sequence terminates,
i.e., $\mathcal{L}^{(n)}_{\rm der}\equiv 0$,  for any $n\ge 5$ in four dimensions.
The list of all  nonzero total derivative terms starting with
the quadratic one reads as,
\beq
\label{L2der0}
\mathcal{L}^{(2)}_{\rm der}(\Pi) &=&[\Pi]^2-[\Pi^2]\,,\\
\label{L3der}
\mathcal{L}^{(3)}_{\rm der}(\Pi)&=&[\Pi]^3-3 [\Pi][\Pi^2]+2[\Pi^3]\,,\\
\label{L4der}
\mathcal{L}^{(4)}_{\rm der}(\Pi)&=&[\Pi]^4-6[\Pi^2][\Pi]^2+8[\Pi^3]
[\Pi]+3[\Pi^2]^2-6[\Pi^4]\,.
\eeq
One  can use the method of
Ref. \cite {drgt} to obtain the two other
polynomials in $\K$ to be included in  massive GR.
For this,  we replace in (\ref{L2der0}-\ref{L4der}) the matrix
elements $\Pi^\mu_\nu$ by the matrix  elements  $\K^\mu_\nu$  defined
in  (\ref {Kmn}). As a result of this procedure, we get the
Lagrangian density \cite {drgt}:
\beq
\mathcal{L}=\frac{\mpl^2}{2} \sqrt{-g} \left ( R + {m^2} (
\mathcal{L}^{(2)}_{\rm der}(\K) + \alpha_3 \mathcal{L}^{(3)}_{\rm der}(\K) +
\alpha_4 \mathcal{L}^{(4)}_{\rm der}(\K) )  \right )\,.
\label{UUU}
\label{explicit}
\eeq
Since all terms in (\ref {Ldern})  with $n\geq 5$
vanish identically,  by construction
all terms $\mathcal{L}^{(n)}_{\rm der}$ with $n\geq 5$
in (\ref {UUU}) are also zero. Hence, the most general Lagrangian density
(\ref {UUU}) has three free parameters, $m,\alpha_3$ and $\alpha_4$.

As it is straightforward to see, Minkowski space is a vacuum with $\phi^a =x^a$,
and the spectrum of the theory  (\ref {UUU})  contains a graviton of mass $m$;
the graviton also has  additional nonlinear interactions specified
by the action at hand.

\subsection{Proof of the absence of FRW cosmologies}

Let us begin by considering  homogeneous and isotropic
solutions to the theory  (\ref {UUU}).  There exists a coordinate system in which
the most general ansatz consistent with these
symmetries reads as follows
\beq
\d s^2=-\d t^2+a^2(t) \d \x^2, \quad \phi ^0 = f(t), \quad  \phi ^i = x^i.
\label{frwans}
\eeq
Here and in the following we are assuming a flat 3-d metric, but our conclusions 
on the absence of the homogeneous and isotropic solutions do not change
if we allow for a more general maximally symmetric 3-space.
Plugging these expressions for the metric and scalar fields into  (\ref {UUU}),
and setting  for  simplicity $\alpha_3=\alpha_4=0$, one obtains the following Lagrangian for $a$ and $f$,
\beq
\mathcal{L}=3 \mpl^2 \( - a \dot a^2-m^2 |\dot f| (a^3-a^2)+ m^2 (2 a^3-3a^2+a) \)   \,,
\label{minisuperL}
\eeq
where overdot denotes the time derivative $\partial_0$. We emphasize that the
quantity  $\dot f$ appears in the Lagrangian
only linearly. The same remains true if we keep  nonzero $\alpha_3$ and $\alpha_4$ - it is just the special
structure  of the terms $\mathcal{L}^{(n)}_{\rm der}(\K),~n=2,3,4$ in (\ref {UUU}),
that  ensures that $\dot f$ enters only linearly!  This is a consequence of the fact that
in the decoupling limit the equations of motion of this  theory have  no more than
two time derivatives acting on the helicity-0 field in particular (and on any field
in general)  \cite {drg2}. Away from the decoupling limit this is
related to the constraint that was found in Refs.
\cite {drgt,hr,drgt_res2}. Here we see the constraint for the FRW metric to all orders, by taking variation of
(\ref {minisuperL}) w.r.t. $f$:
\beq
\label{frwconstr}
m^2 \partial_0 (a^3-a^2)=0 \,.
\eeq
This constraint makes time evolution of the scale factor impossible.
As we have noted above,  keeping the $\mathcal{K} ^3$ and $\mathcal{K}^4$ terms in  (\ref{UUU})
can only modify the polynomial  function of $a$ on which $\partial_0$ acts in  (\ref{frwconstr}).
Therefore,  there are no nontrivial homogeneous and isotropic solutions in
the theory of massive GR, defined by (\ref{UUU}).

It is also instructive to  show   the absence of FRW solutions in the unitary gauge, for which
$\phi^a=\delta^a_\mu x^\mu$, and no $f$ field appears in the action to begin with.
In this gauge, the most general homogeneous and isotropic ansatz
involves the lapse function $N(t)$,
\beq
\d s^2=-N^2(t) \d t^2+a^2(t) \d  \x^2\,,
\eeq
and the Lagrangian  (\ref{UUU}) with $\alpha_3=\alpha_4=0$  reads
\beq
\mathcal{L}=3\mpl^2 \(-\frac{a \dot a^2}{N}-m^2 (a^3-a^2)+ m^2 N (2 a^3-3a^2+a) \).
\label{larg11}
\eeq
As can be straightforwardly verified, the condition (\ref{frwconstr}) in this case arises as the
requirement of  consistency of the equations of motion for the two fields, $a$ and $N$
in (\ref{larg11}). More specifically, one can obtain (\ref{frwconstr}) by taking the difference between
the time-derivative of the e.o.m.  for $N$ and the e.o.m. for $a$. Technically, this is so because
the second term on the r.h.s. of (\ref {larg11}) has no factors of $N$ in it and the constraint
arises as  the direct result of the Bianchi identity of GR.

We briefly note that the homogeneous and isotropic solutions would not be forbidden
if the mass term were  not an explicit  constant,  but instead emerged
as a VEV of some field-dependent function; i.e., if we replaced $m^2 \to m^2(\sigma)$  in
(\ref{UUU}), where $\sigma$ is a scalar field that also has its own  kinetic and  potential terms.
Then, variation  w.r.t. $f$ would give rise to a
constraint
\beq
\partial_0(m^2(\sigma) (a^3-a))=0\,,
\label{sigmcon}
\eeq
that  relates time evolution of the scale factor to that of the $\sigma$ field,
but it does not forbid  homogeneous and isotropic solutions. Hence, the absence of the
homogeneous and isotropic solutions is an intrinsic property of massive GR with an
explicit mass term, as in (\ref{UUU}). By this property it could potentially be
distinguished observationally from the theory with a dynamical mass $m^2(\sigma)$.
Moreover, for the latter theory one should expect the presence of an additional massless 
(or very light) scalar at cosmological distances, which is hidden by the Vainshtein 
mechanism at shorter scales.
One example of this is when $m^2 \to m^2{\rm exp }(\sigma/\Mpl)$, for which the  DL theory (with
the kinetic term for $\sigma$)  reduces to a theory with two galileons coupled to the tensor field.

\vspace{0.1in}

The above-described properties of massive GR  are similar to those of
a peculiar scalar field theory (the so-called  Cuscuton), defined by the following Lagrangian~\cite {cuscuton}:
\beq
\mathcal{L}=\mu^2\sqrt{-g}\sqrt{|g^{\mu \nu}\p_\mu\phi\p_\nu\phi|},
\label{Cus}
\eeq
where $\mu$ is some  dimensionful constant.  Assuming  $\phi=\phi(t)$,  and the
homogeneous and isotropic FRW metric  (\ref{frwans}), the scalar field equation
reduces to a constraint, similar to (\ref{frwconstr})
\beq
\p_\mu\left (\sqrt{-g}\frac{g^{\mu \nu}\p_\nu\phi}{\sqrt{|g^{\alpha\beta}\p_\alpha\phi\p_\beta\phi|}}\right )=0~
\Rightarrow \p_0a^3 =0.
\eeq
Therefore, the theory (\ref {Cus})  does not possess homogeneous and isotropic
cosmological solutions in full analogy to massive GR  described above.  Pursuing this analogy further,
the homogeneous and isotropic cosmological solutions  would be permitted if we were to promote
the parameter $\mu$ into a field dependent function $\mu^2 \to \mu^2(\sigma)$. In this case
the constraint would read as $\p_0 (\mu^2 (\sigma) a^3)=0$; the latter links the time
evolution of the scale factor to that of $\sigma$,  but it does not
forbid  homogeneous and isotropic cosmological solutions.

We note that the equations of motion of the theory (\ref {Cus}) are invariant under the
replacement $\phi \to E(\phi)$, where $E$ is an arbitrary differentiable function. Hence, by finding
a particular solution one immediately generates an infinite number of solutions.
Whether a somewhat similar invariance exists in the equations of motion of
massive GR is not obvious.

Last but not least, we complement the present section with the discussion of the degravitating solution of
massive GR (for degravitation see \cite {degrav}, for the corresponding decoupling-limit solution, 
see \cite{drghp}). Although the
constraint (\ref{frwconstr}) forbids the time-evolving FRW solutions, it still allows for a static Minkowski
metric - even in the presence of a Cosmological Constant. To see this, we note that the action
(\ref {larg11}) in the presence of the vacuum energy density  ${\cal E}$ has an additional contribution
of the form $-N a^3 {\cal E}$. In this case one can determine the value of the scale factor
from the $N$-equation, which for a static homogeneous field reduces to,
\beq
2a^3-3a^2 +a =  a^3 \frac{{\cal E}}{\rho_{\rm co}},
\eeq
where, as before, $\rho_{\rm co}= 3\mpl^2 m^2$. The  value of $N$ is determined from the
equation of motion for $a$. The solution to this equation exists if the following
inequality is satisfied,
 \beq
 {\cal E} \geq - \frac{\rho_{\rm co}}{4•}\,.
 \eeq
Hence, the degravitation works for an arbitrarily large
positive vacuum energy density. If we were to include also an arbitrary $\alpha_3$ and $\alpha_4$,
degravitation could have been  achieved for arbitrary ${\cal E}$.  However, we note that
fluctuations on the degravitated background  exhibit the Vainshtein mechanism
at the scale determined by the degravitated vacuum energy --   the larger the degravitated
energy, the smaller is the corresponding Vainshtein radius \cite {drghp}. Because of this, there is an unscreened
fifth force, and we cannot be living today on such a background. Hence, the degravitation could  have
only taken place  in a far past after which the universe must have  transitioned to a different background
(see more in \cite {drghp}). Note that the screening solution exists for a broad class of external sources, not
just for a Cosmological Constant.

\section{Cosmology at high densities: $\rho \gg \rho_{\rm co}$}

We begin by thinking of the universe filled with pressure-less dust of  density $\rho \gg \rho_{\rm co}$, with $\rho_{\rm co}$ defined in (\ref {rhoco}).  As was discussed in Section 1,  cosmological evolution of such a universe  can  very well be approximated by  the standard  FRW metric of GR. This is so because an arbitrary Hubble patch enclosed by a sphere of radius $H^{-1}= (\rho/3\mpl^2)^{-1/2}$, is well within its Vainshtein radius, $H^{-1}\ll r_*$. Hence, the early universe in massive GR  would  evolve as it  does in GR, with some  small corrections.
These corrections,  for any observer in such a universe, can be estimated as some positive
power  of the ratio $(m/H)\ll 1$.  On the other hand, we would expect the scalar fields, $\phi^a$,
to be in a nonperturbative (Vainshtein) regime at these scales, and yet, their stress-tensor should be
sub-dominant to the matter/radiation stress-tensor that drives the FRW expansion.
That this is so is a necessary condition for the self-consistency of the
solution.  Below, we will calculate the expressions for the scalar fields $\phi^a$
in an FRW background, and  discuss its backreaction.

The most general spherically symmetric solution,  including the four scalars $\phi^a$,
can always be put in the following form,
\begin{gather}
\rmd s^2 = - \rmd t^2 + C(t,r) \rmd t \rmd r + A^2(t,r) \left[\rmd r^2 + r^2 \rmd \Om^2 \right] \; , \nonumber \\
\phi^0 = f(t,r) \; , \qquad \; \phi^i = g(t,r) \frac{x^i}{r} \; .
\label{eq:spherical}
\end{gather}
The advantage of the latter form of the metric is that it is easier to compare to the standard FRW,
while the appropriate $\phi^a$ fields can be treated separately.

The Einstein equation, obtained by varying (\ref{UUU}) with $\alpha_3=\alpha_4=0$ with respect to the metric, reads as follows,
\be
\label{eq:einst}
G_{\mu \nu} = m^2 T^{(K)}_{\mu \nu} + \frac{1}{ \Mpl^2} T^{(m)}_{\mu \nu} \; ,
\ee
where $T^{(K)}_{\mu \nu} $ is the effective stress tensor due to the mass term
in (\ref {UUU}), while  $T^{(m)}_{\mu \nu}$ denotes the stress energy tensor of
standard matter.  Taking a covariant derivative of the above equation leads to
the Bianchi constraint, $m^2   \nabla^\mu T^{(K)}_{\mu \nu}=0, $ which is just the
equation of motion obtained by varying the action w.r.t. $\phi^a$,
\be
\label{eq:constraint'}
\frac{\del S}{\del \phi^a}\,=0 \,.
\ee
As discussed above, we will be neglecting at the zeroth order the $m^2 T^{(K)}_{\mu \nu}$ term, as well as the (possible) $r$ dependence in the metric \eqref{eq:spherical}.
Hence, the zeroth order solution for the metric will be the standard FRW solution, with $A(r,t)=a(t)$ and $C(t,r)=0$ corresponding to
the matter content encoded in $T^{(m)}_{\mu \nu}$.
For the scalars $\phi^a$, instead, we have to solve the full equations \eqref{eq:constraint'}
in the background FRW metric just defined, since these are already proportional to $m^2$.
For this, we can rewrite the potential part of the massive GR action (\ref{L2K}) using the following identity,
\beq
\K^\mu_\nu \K^\nu_\mu -(\K^\alpha_\alpha)^2=-12+ 6 ~\operatorname{tr}\sqrt{g^{-1}\Sigma}+\operatorname{tr} g^{-1}\Sigma-(\operatorname{tr}\sqrt{g^{-1}\Sigma})^2,
\eeq
where the matrix $(g^{-1}\Sigma)^{\mu}_{~\nu}$ is defined as follows:
\beq
(g^{-1} \Sigma )^{\mu}_{~\nu} = g^{\mu\alpha}\p_{\alpha}\phi^a \p_{\nu}\phi^b\eta_{ab}\equiv g^{\mu\alpha}\Sigma_{\alpha\nu}=
\begin{pmatrix}
\dot{f}^2 - \dot{g}^2 & \dot{f} f' - \dot{g} g' & 0 & 0 \\
\frac{\dot{g} g' - \dot{f} f'}{a^2} & \frac{-f'^2 + g'^2}{a^2} & 0 & 0 \\
0 & 0 & \frac{g^2}{a^2 r^2} & 0 \\
0 & 0 & 0 & \frac{g^2}{a^2 r^2}
\end{pmatrix}~.
\eeq
Evaluating the eigenvalues of the latter matrix, and varying the resulting action w.r.t the two fields $f(t,r)$ and $g(t,r)$, one obtains the corresponding equations of motion,
\be
\label{eq:const1}
\de_t \left[ \frac{y}{\sqrt{X}} \left(\dot{f} + \mu \frac{g'}{a} \right) + \mu a^2 r^2 g' \right]
- \de_r \left[ \frac{y}{a \sqrt{X}} \left(\frac{f'}{a} + \mu \dot{g} \right) + \mu a^2 r^2 \dot{g} \right] = 0\,,
\ee
\begin{multline}
\label{eq:const2}
\de_t \left[ \frac{y}{\sqrt{X}} \left(\dot{g} + \mu \frac{f'}{a} \right) + \mu a^2 r^2 f' \right]
- \de_r \left[ \frac{y}{a \sqrt{X}} \left(\frac{g'}{a} + \mu \dot{f} \right) + \mu a^2 r^2 \dot{f} \right] \\
- 6 a^2 r + 2 a g + 2 a^2 r \sqrt{X} = 0 \; ,
\end{multline}
where dot denotes time derivative and prime denotes derivative w.r.t. $r$, and  we have
introduced the following notation,
\be
y \equiv 2 g a^2 r - 3 a^3 r^2 \; , \quad
X \equiv \left(\dot{f} + \mu \frac{g'}{a} \right)^2 - \left(\dot{g} + \mu \frac{f'}{a} \right)^2
\;, \quad
\mu \equiv \mathrm{sgn} (\dot{f} g' - \dot{g} f') \; .
\label{defs}
\ee

By solving (\ref{eq:const1})-(\ref{eq:const2}) for the \stu fields, and calculating the effective
stress-energy tensor $T^{(\K)}\mn$ on the solution, one can evaluate the backreaction on the geometry
from the presence of $\phi^a$'s. For the Vainshtein mechanism to be operative, two conditions should
be met inside the Vainshtein radius: (I) the backreaction of
the  \stu  stress-tensor should be negligible, so that the background evolution is
described by the FRW geometry to a very high precision, and  (II) the metric fluctuations should be those
of GR to a very high precision. This would guarantee, that the scalar is successfully screened.
Generically  one would expect these properties to hold in the Vainshtein region
as the stress-tensor of the \stu fields in multiplied
by a small parameter $m^2$. Hence, one should anticipate
significant departures for the standard  GR results at scales of order $1/m$.
Even though the metric is homogeneous and isotropic, the \stu fields
are not. Due to these inhomogeneous fields  there is  a physical center in each $1/m$-size 
domain. This center is not felt by matter coupled to the metric, but
perturbations will be sensitive to it. This by itself  restricts the value of
$m$ to be an order of magnitude smaller than $H_0$,  or  less.

For illustrative  purposes, we give  in Appendix A  a particular exact solution, which
satisfies the condition (I) in an interesting way.  On that solution   the stress-tensor
$T^{(\K)}\mn$ \textit{exactly} coincides with the stress-tensor of
dark energy with the energy density $\sim \rho_{co}$, in spite of the fact that
the \stu fields are inhomogeneous! As a result, this solution can exist even when 
no external stress-tensor  is introduced. Hence, it is in a class of self-accelerated solutions.
This solution  can  exist not only in the Vainshtein regime, but also outside of it. 

What is however not clear is  whether the fluctuations on this self-accelerated solution are
close to those of GR (it is easy to show that subhorizon fluctuations are, but one needs to demonstrate 
it for larger scales as well, which requires some careful calculations).  
Until this is known  the solution can only serve a
demonstrational purpose showing  the smallness of the backreaction in the Vainshtein regime.
This is precisely how we regard this solution in the present work.

Putting the issues  of perturbations aside,  on the self-accelerated
solution the value of the mass is related to that of the present-day Hubble parameter
as $m^2 = C\,H_0^2$, where $C$ is a free constant in the theory, which would
depend on the parameters $\alpha_3$ and $\alpha_4$ (if we were to include them); 
without significant tunings of these  parameters the theoretical 
value of $C$ should not be assumed to be  outside of the interval $C\sim (0.01-1)$.
In this case, it should be possible to rule out such a scenario observationally (or at least 
to rule out a significant fraction of the parameter space for $\alpha_3,\alpha_4$), as at the 
present-day Hubble scales one would expect departures from the FRW evolution 
of the order of $C$.

\section{Cosmology at low densities: $\rho \ll \rho_{\rm co}$ }
\label{sec:FPcosmo}

After the energy density $\rho$ drops below its crossover value (\ref {rhoco}),
massive gravity enters the linear  regime, as discussed in Section 1. In this regime, no
matter how small the graviton mass, the  massive theory differs from the massless one
by quantities  of order ${\cal O}(m^0)$, thus  exhibiting the vDVZ discontinuity \cite {vdvz}.
Therefore, the cosmology described by the massive theory is expected to differ significantly
from the conventional one.  The purpose of this section is to study that cosmology.
For this we first recall the status of linearized cosmology
in GR. There are some subtleties in this, and we would like to emphasize those
relevant for our  discussions.

First, the  matter/radiation stress-tensor should be conserved
in this approximation,  $\partial^\mu T_{\mu\nu}=0$.
Then, if we were to choose a diagonal stress-tensor, $T^{\mu}_{\nu}={\rm diag} (-\rho, p,p,p)$,
the conservation would impose  $\rho$  to be  time-independent.
To avoid this restriction, we will have to choose a
coordinate system in  which the stress-tensor is not diagonal and takes the form:
\beq
 T_{\mu\nu} =
 \begin{pmatrix}
  \rho & -H(\rho+p)x^i \\
  -H(\rho+p)x^i & p\delta_{ij}    \\
  \end{pmatrix},
\label{stress}
\eeq
where $H,\rho,p$ are arbitrary time-dependent functions.
It is straightforward to check then that
the condition  $\partial^\mu T_{\mu0}=0$ leads to the  proper  conservation
equation,  ${\dot \rho} + 3H (\rho +p)=0$, as soon as
$H,\rho$ and $p$ are interpreted as the Hubble parameter, density and pressure respectively
\footnote{For momentum conservation one should use the full covariant expression
$\nabla^\mu T_{\mu i}$ that leads to the acceleration equation;
in this case however, it is obtained
at the \textit{linear} order in $\vec x$ and requires correcting the stress-tensor
by $\mathcal{O} (H^2\bold x^2)$ quantities~\cite{guido}.}.

Second, strictly speaking, linearized GR itself has no homogeneous and isotropic cosmology:
indeed, assuming that $h_{\mu\nu}$ is a function of $t$ only,  the $00$ component of the Einstein
tensor vanishes, and the   $00$ component of the Einstein equation cannot be satisfied.
The way out is to resort to the Fermi coordinate system in which the metric takes an inhomogeneous form
(for recent discussions see, e.g., \cite{baumann}):
\beq
\d s^2=-\left ( 1-(\dot H+H^2)\bold{x}^2 \right ) \d t^2+
\left( 1-\frac{1}{2}H^2 \bold{x}^2\right)\d\bold{x}^2=
\left(\eta_{\mu\nu}+h^{\rm FRW}_{\mu\nu}\right)\d x^\mu \d x^\nu
\label{FRW},
\eeq
where the corrections to the above expression are suppressed by higher powers of
$H^2\bold{x}^2$.  As long as $H^2\bold{x}^2\ll 1$ , the above metric describes cosmology
in any local patch as a small deviation from
Minkowski space.

In GR there exists a coordinate transformation
that brings (\ref {FRW}), in the approximation considered,  to a homogeneous and
isotropic metric in a  comoving coordinate system $t_c,
{\vec x}_c $,  in which the stress-tensor has a conventional form, $T_{\mu\nu}=
{\rm diag}(\rho(t_c), \delta_{ij} a^2(t_c)p(t_c))$. This coordinate transformation
takes the form (see \cite {nrt})
\beq
t_c= t - {1\over 2}H(t) {\vec x}^2,~~~{\vec x}_c = {{\vec x}\over a(t)} \left (1+ {1\over 4}H^2(t)
{\vec x}^2   \right )\, ~,
\label{ttc}
\eeq
however, the transformation itself is essentially
nonlinear\footnote{This is consistent with the fact that the Friedmann equation in the comoving system
relates the {\it square} of the perturbation to density, $(\dot {\delta a})^2\sim G_N \rho$,
where $\delta a$ denotes a departure of the scale factor from the Minkowski space, $\delta a = a-1$.}.

\vspace{0.1in}

Linearized massive gravity is not much different in that respect -- it does not admit
homogeneous and isotropic solutions either. We  prove this by assuming the opposite
and showing the contradiction. For this, consider the
Fierz-Pauli (FP) Lagrangian to which any consistent Lorentz-invariant
massive gravity should reduce at the linearized  level \cite{fp} ,
\beq
\mathcal{L}=-\frac{1}{2}h^{\mu\nu}
\mathcal{E}_{\mu\nu}^{\rho\sigma}h_{\rho\sigma}-\frac{1}{4}m^2(h^{\mu\nu}h_{\mu\nu}-h^2)+
h^{\mu\nu}T_{\mu\nu} \, .
\label{fp}
\eeq
Here, $h\mn\equiv g\mn-\eta\mn$ and $\mathcal{E}$ denotes the linearized Einstein operator
\beq
\mathcal{E}_{\mu\nu}^{\rho\sigma}h_{\rho\sigma}=- {1\over 2} ( \square h_{\mu \nu} - \partial_\mu \partial^\alpha
h_{\alpha\nu} - \partial_\nu \partial^\alpha
h_{\alpha\mu} + \partial_\mu \partial_\nu h -
\eta_{\mu\nu} \square h + \eta_{\mu\nu}
\partial_\alpha  \partial_\beta h^{\alpha\beta}) \, ,
\label{Eterm}
\eeq
all indices are contracted with the flat metric,
and the Planck mass has been set to one.  Furthermore, applying $\p^\mu$ to the
equation of motion obtained from (\ref {fp}) and using the Bianchi identity,
one gets a  constraint,
\beq
\p^\mu h_{\mu\nu}=\p_\nu h.
\label{Bianchi}
\eeq
In  the unitary gauge the  metric perturbation $h_{\mu\nu}$ represents a \textit{physical}
field, and  requiring homogeneity  makes all of its components space-independent,
\beq
\p_{i}h_{\mu\nu}=0.
\eeq
With this assumption, the constraint  (\ref{Bianchi}) reduces to the following equations
\beq
h_{0i}=\textit{const}\equiv c_i, \quad h_{ii}=\textit{const}\equiv c.
\label{constr}
\eeq
The second of these equations implies that either a solution is trivial or else it
cannot be isotropic. Hence, linearized massive GR has no nontrivial
homogeneous and isotropic solutions. Moreover, it is straightforward to
check that in general, inhomogeneity cannot be removed entirely into the longitudinal degrees of freedom.

\vspace{0.1in}

Then, what is a  snapshot of the universe at $\rho \ll \rho_{\rm co}$?
We will show below that it can be pictured  as a collection of
multiple domains, each of size $1/m$,  such that  well within a given domain,
at scales $ \ll 1/m$,  cosmology deviates significantly from the conventional GR one. 
How these domains  are glued together is a complicated question that is not addressed here.
Nevertheless, at a scale  much greater than  $1/m$, when averaged over many domains
enclosed by this scale, the universe should look  homogeneous
and isotropic again.

What exactly is  then a  solution to the linearized massive GR  in each domain?
To get this solution, it is useful to introduce appropriately
normalized  St\"uckelberg fields:
\beq
h_{\mu\nu} = {\bar h}_{\mu\nu} + \p_\mu V_\nu +\p_\nu V_\mu = {\bar h}_{\mu\nu} +
{\p_\mu A_\nu +\p_\nu A_\mu\over m} + {2 \p_\mu\p_\nu \pi\over m^2}\,,
\label{stu}
\eeq
and consider scales that are much smaller than  $1/m$.
In this  approximation, the FP theory (excluding the totally decoupled vector mode) reduces to:
\beq
\mathcal{L}=-\frac{1}{2}{\bar h}^{\mu\nu}
\mathcal{E}_{\mu\nu}^{\rho\sigma}{\bar h}_{\rho\sigma}-
{\bar h}^{\mu\nu} (\p_\mu \p_\nu \pi - \eta_{\mu\nu}\square \pi)
+{\bar h}^{\mu\nu}T_{\mu\nu} + {\cal O}(m^2)\, .
\label{fp1}
\eeq
The latter can be diagonalized by the conformal shift
${\bar h}_{\mu\nu}=  {\tilde h}_{\mu\nu}+\eta_{\mu\nu} \pi $,
giving rise to the Lagrangian
\beq
\mathcal{L}=-\frac{1}{2}{\tilde  h}^{\mu\nu}
\mathcal{E}_{\mu\nu}^{\rho\sigma}{\tilde  h}_{\rho\sigma} + {3\over 2} \pi \square \pi
+{\tilde  h}^{\mu\nu}T_{\mu\nu}+  \pi T  + {\cal O}(m^2)\, .
\label{fp2}
\eeq
The equations of motion that follow from this Lagrangian are:
the GR equations for ${\tilde h}_{\mu\nu}$, and a simple equation for
$\pi$,  $\square \pi = -T/3$. The solution for
${\tilde h}_{\mu\nu}$, as argued above, is  ${h}^{FRW}_{\mu\nu}$  given in (\ref {FRW}),
while for $\pi$ we obtain ${\square \pi_{\rm sol}} =2 ({\dot H} + 2 H^2)$.
As a result, the physical metric is ${h}^{Phys}_{\mu\nu} ={h}^{FRW}_{\mu\nu}+ \eta_{\mu\nu} \pi_{\rm sol}$,
and the interval takes the form:
\beq
\d s^2=-\left ( 1-{1\over 3} (2 \dot H+H^2)\bold{x}^2 \right ) \d t^2+
\left( 1+ \frac{1}{6}( 2 \dot H+H^2) \bold{x}^2\right)\d\bold{x}^2\,.
\label{phys}
\eeq
Note that the linearized Ricci curvature on the physical metric
${h}^{Phys}_{\mu\nu}$ is zero (in the leading approximation), as it should be the 
case for all cosmologies due to (\ref {Bianchi}).  This metric  deviates from  its GR counterpart by the value
of $\pi_{\rm sol}$, which is of the same order as ${h}^{FRW}_{\mu\nu}$ itself.
This is a cosmological manifestation of the vDVZ discontinuity.

At scales $\sim 1/m$  the mass terms neglected in (\ref {fp2})
should be reinstated and used.  However, at yet larger scales,
which enclose a large  number of domains, the universe  should look  homogeneous,
if we average over all enclosed domains.
We notice that at scales $\gg 1/m$, all the derivative terms in the
Lagrangian (\ref {fp}) should be neglected and only the mass terms
should be kept. Then, the equation of motion takes the form:
\beq
m^2 (h_{\mu\nu} -\eta_{\mu\nu}h) = 2 \langle T_{\mu\nu} \rangle \,,
\label{meq}
\eeq
where  $\langle T_{\mu\nu} \rangle $  is  the stress-tensor (\ref {stress}) averaged over many
$1/m$-size domains. At such large scales gravity is screened.  Depending on 
the initial conditions there may be a number of different scenaria of how
one could match the large and small scale  behavior to each other.
If the $1/m$-size domains are densely packed and adjacent to each other, 
then the matter/radiation  will get   accumulated  near the 
boundaries of the domains,  as it is expelled from the bulk.  However,
if the domains are  well separated, or there is only one domain, then density near the 
boundaries  should be expected to be suppressed since  gravity 
of the bulk material is screened and matter particles will
free stream  out of the domain. All these scenaria, and the initial conditions that could 
give rise to them, need separate detailed studies.

\subsection*{Acknowledgments}
We would like to thank Matt Kleban and Roman Scoccimarro for useful comments. The work of GD'A is supported 
by the James Arthur Fellowship. CdR is supported by the Swiss National Science Foundation. 
The work of SD is supported in part by the NSF grant PHY-1068438. SD thanks the hospitality 
of the  Kavli Institute for Theoretical Physics at UC Santa Barbara (under the NSF grant PHY05-51164), 
whilst this work was being completed. GG is supported by the NSF grant  PHY-0758032. 
The work of DP  is supported by Mark Leslie Graduate Assistantship. AJT would like to 
thank the Universit\'e de  Gen\`eve for hospitality whilst this work was being completed.

%\newpage

\renewcommand{\theequation}{\Roman{equation}}
\setcounter{equation}{0}

\appendix

\section{An exact background solution}

We will now show that it is possible to find a solution of the system (\ref{eq:const1})-(\ref{eq:const2})
\be
\label{eq:phi1}
\de_t \left[ \frac{y}{\sqrt{X}} \left(\dot{f} + \mu \frac{g'}{a} \right) + \mu a^2 r^2 g' \right]
- \de_r \left[ \frac{y}{a \sqrt{X}} \left(\frac{f'}{a} + \mu \dot{g} \right) + \mu a^2 r^2 \dot{g} \right] = 0\,,
\ee
\begin{multline}
\label{eq:phi2}
\de_t \left[ \frac{y}{\sqrt{X}} \left(\dot{g} + \mu \frac{f'}{a} \right) + \mu a^2 r^2 f' \right]
- \de_r \left[ \frac{y}{a \sqrt{X}} \left(\frac{g'}{a} + \mu \dot{f} \right) + \mu a^2 r^2 \dot{f} \right] \\
- 6 a^2 r + 2 a g + 2 a^2 r \sqrt{X} = 0 \; .
\end{multline}

We notice that \eqref{eq:phi1} is identically satisfied for $y = 0$, or
\be
g(t,r) = {3\over 2} a(t) r \; .
\label{eq:gsol}
\ee
For this ansatz  eq. \eqref{eq:phi2} reduces to the following equation for $f(t,r)$:
\beq
\sqrt{\left( \mu \dot{f} + \frac{3}{2} \right)^2 - \left(\frac{3}{2} H a r + \mu \frac{f'}{a} \right)^2 }=\mu \dot f +\frac{3}{2}- \mu H r f' \,,
\label{eq:feq}
\eeq
with $H=\dot a/ a$ denoting the usual Hubble parameter.

The structure of the equation~\eqref{eq:feq} guarantees that all square roots appearing in it are always well-defined. Indeed, the above equation for $f$ can be reduced to a simpler one by rewriting it as follows,
\beq
\sqrt{ \left( \mu \dot{f} + \frac{3}{2} -\mu H r f' \right )^2-\frac{9}{4•}H^2 a^2 r^2-\frac{f'^2}{a^2•}+2 H r \dot f f'-H^2r^2f'^2 }= \nn \\ \mu \dot f +\frac{3}{2•}- \mu H r f'. \label{fneq}
\eeq
Squaring the latter equation puts it in the following form:
\beq
\frac{f'^2}{a^2} \( 1 + a^2 H^2 r^2\) - 2 H r \dot f f'  + \frac{9}{4} a^2 H^2 r^2 =0 \, .
\label{eq:fsimple}
\eeq
Obviously, the expression under the square root in (\ref{fneq}) is positive semi-definite for any solution of the squared equation. Moreover, for any solution of this equation, the quantity appearing on the r.h.s of (\ref{fneq}) is positive, which can be seen from writing it as $$  \mu \dot f +\frac{3}{2•}- \mu H r f' = \frac{2}{3 a•}\mu\(\dot f g'-\dot g f' \)+\frac{3}{2•}~,$$ and recalling the definition of $\mu$, eq. \eqref{defs}.
Therefore, any solution of eq. \eqref{eq:fsimple} will solve eq. \eqref{eq:feq}.

One solution of eq. \eqref{eq:fsimple} is
\beq
f(t,r) = \frac{9}{16 T} \int^t \frac{\rmd \tilde{t}}{a(\tilde{t}) H(\tilde{t})} + a(t) T \( 1 + \frac{9 r^2}{16 T^2} \) \, ,
\label{eq:fsol}
\eeq
where $T$ is an integration constant with dimensions of time, and the choice of the lower limit of integration corresponds to a constant shift of $\phi^0$.

We are now in a position to  compute  the stress-tensor of the $\phi^a$ fields and compare it with that of matter.
It follows from (\ref{L2K}), that the effective stress-tensor is given by the following expression:
\begin{align}
T^{(K)}_{\mu \nu}& = \frac{1}{\sqrt{-g}} \frac{\del}{\del g^{\mu \nu}}
\left[ \sqrt{-g} \left(  \mathcal{K}^2_{\alpha\beta}- \mathcal{K}^2  \right) \right] \nn \\
 &= \frac{1}{•2} \bigg [ \( 12 + (\Tr \sqrt{g^{-1} \Sigma} - 6) \Tr \sqrt{g^{-1} \Sigma} - g^{\alpha \beta} \Sigma_{\alpha \beta} \) g_{\mu \nu} \nn \\
&+ 2 \Sigma_{\mu \nu}
+ (3-\Tr \sqrt{g^{-1} \Sigma}) \( g_{\mu \al} \big(\sqrt{g^{-1} \Sigma}\big)^\al_{\; \nu} + g_{\nu \al} \big(\sqrt{g^{-1} \Sigma}\big)^\al_{\; \mu} \) \bigg ]  ~.
\end{align}
The nonzero components of $T^{(\K)}\mn$ are therefore:
\begin{align*}
 T^{(K)}_{00} = \frac{1}{•2} \bigg [ -12-4 \sqrt{X} \( \frac{g}{ar•}-\frac{3}{2•} \)-2\frac{g}{ar•} \( \frac{g}{ar•}-6 \)
 \nn \\ +\frac{2}{\sqrt{X}•} \( 3-2\frac{g}{ar•} \) \bigg ( \dot g^2-\dot f^2 -\frac{\mu}{a•}\( \dot f g' - \dot g f' \) \bigg ) \bigg ] ~,
\end{align*}
\begin{align*}
T^{(K)}_{0r} =\frac{1}{\sqrt{X}•}\( 3 -2\frac{g}{ar•}\)\( \dot g g'-\dot f f' \)~,
\end{align*}
\begin{align*}
 T^{(K)}_{rr} = \frac{1}{•2}  a^2 \bigg [ 12+4 \sqrt{X} \( \frac{g}{ar•}-\frac{3}{2•} \)+2\frac{g}{ar•} \( \frac{g}{ar•}-6 \)
 \nn \\ +\frac{2}{\sqrt{X}•} \( 3-2\frac{g}{ar•} \) \bigg ( \(\frac{g'}{a}\)^2-\(\frac{f'}{a}\)^2 +\frac{\mu}{a•}\( \dot f g' - \dot g f' \) \bigg ) \bigg ] ~,
\end{align*}
and
\begin{align*}
 T^{(K)}_{\theta\theta} =\frac{T^{(K)}_{\phi\phi}}{\sin ^2 \theta}= \frac{1}{•2} a^2 r^2 \bigg [ 12+4 \sqrt{X} \( \frac{g}{ar}-\frac{3}{2} \)+2\frac{g}{ar} \( \frac{g}{ar}-6 \)
 \nn \\ + 2\frac{\mu}{a}\(\dot f g'-\dot g f'\)+ 2 \( \frac{g}{ar} \)^2+2 \frac{g}{ar•}\(3-\sqrt{X}-2\frac{g}{ar}\)   \bigg ]~ .
\end{align*}
Remarkably enough, as we show below, for the solution at hand the inhomogeneities of the St\"uckelberg fields completely fall out from the expression for the effective stress-tensor $T^{(K)}\mn$.

Using the exact solution for $g(r,t)$, as well as the equation of motion for $f(r,t)$ \eqref{eq:fsimple}, $T^{(\K)}\mn$ \emph{exactly} reduces to the diagonal, cosmological-constant-type form - with the corresponding Hubble scale set by the value of the graviton mass,
\beq
{T^{(\K)}}^{\mu}_{~\nu}= \begin{pmatrix}
-\rho & 0& 0 & 0 \\
0 &p& 0 & 0 \\
0 & 0 &p & 0 \\
0 & 0 & 0 &p
\end{pmatrix},\quad \rho=-p=\frac{3}{4} m^2 \mpl^2 =\frac{1}{4}\rho_{co}~.
\eeq
Therefore, the backreaction from the \stu fields is indeed negligible for the universe filled with matter or radiation with density significantly exceeding $\sim \rho_{\rm co}$.

The background solution we just found, consisting of the FRW metric plus eqs. \eqref{eq:gsol} and \eqref{eq:fsol}, is exact.
Cosmology for this solution - at least at the background level - is therefore completely insensitive to the presence of inhomogeneity in the \stu scalars.
Even in the absence of any external sources, the geometry describes the homogeneous and isotropic self-acceleration of the universe with the Hubble constant equal to $m/2$, all of the inhomogeneities of the solution being removed into the \stu scalars.
On the other hand, the space-dependence of the background value of $\phi^a$ can be probed by perturbations on the FRW metric.

More generally, we expect the theory to admit solutions with truly inhomogeneous geometry - with the metric being impossible to put in a homogeneous form by any coordinate transformations.
For such solutions, we expect the backreaction of the \stu fields to be negligible in the high-density regime, while becoming important for densities $\lsim  \rho_{\rm co}$.
We emphasize again that the question of whether the evolution can continue on the self-accelerated  solution, or alternatively should switch to the low density vDVZ regime, remains open, as the matching between these two regimes is hard to study analytically.

\section{Anisotropic solutions}

In this Appendix, we show that there are cosmological solutions which are homogeneous but  anisotropic.

\subsubsection*{Parity symmetric solutions}

Let us start by considering the following diagonal Ansatz for the unitary gauge metric:
\be
\rmd s^2 = - N^2(\tau) \rmd \tau^2 + a_1^2(\tau) \rmd x^2 + a_2^2(\tau) \rmd y^2 + a_3^2(\tau) \rmd z^2 \; ,
\ee
which is the most general homogeneous metric invariant under the discrete parity symmetry $\x \to - \x$.
Before plugging this into the action, it is convenient to redefine the time variable as
\be
\rmd t = N(\tau) \rmd \tau \; .
\ee
This gauge transformation will excite one of the \stu fields, which will now read:
\be
\phi^0 = f(t) \; , \qquad \phi^i = x^i \; .
\ee
In this gauge we can immediately derive a constraint on the scale factors.

The matrix $\Sigma^\mu_{\; \nu} \equiv g^{\mu \al} \de_\al \phi^a \de_\nu \phi_a$ is diagonal, with the following eigenvalues:
\be
\sig_0 = \dot{f}^2 \; , \quad \sig_1 = \frac{1}{a_1^2} \; , \quad \sig_2 = \frac{1}{a_2^2} \; , \quad \sig_3 = \frac{1}{a_3^2} \; .
\ee
The mass term in the action therefore reads
\begin{multline}
S = \frac{\Mpl^2}{2} m^2 \int \rmd^4 x \bigg[ - 12 a_1 a_2 a_3 + 6 (a_1 a_2 + a_1 a_3 + a_2 a_3) - 2 (a_1 + a_2 + a_3) \\
+ 2 \dot{f} (3 a_1 a_2 a_3 - a_1 a_2 - a_1 a_3 - a_2 a_3) \bigg] \; .
\end{multline}
It is immediate to derive the following algebraic constraint on the scale factors, which is proportional to $m^2$ and it is valid independently of the matter Lagrangian coupled to the metric:
\be
\label{eq:constraint}
3 a_1 a_2 a_3 - (a_1 a_2 + a_1 a_3 + a_2 a_3) = k \; ,
\ee
where $k$ is an integration constant.
As a check of the calculations, we can easily see that there is no isotropic solution other than Minkowski space-time; for $a_1 = a_2 = a_3 = a$, we have $a^3 - a^2 = k$, which is exactly eq. \eqref{frwconstr}.

It is simple to show that we cannot have a solution for which all the scale factors grow at small times.
In fact, requiring that $a_i \to 0$ as $t \to 0$, we see that we should set $k = 0$, because the whole LHS 
should vanish for small times. However, since the scale factors are positive, the equation cannot be satisfied.
Thus, we cannot have a solution with all the scale factors growing with time.
The only non-trivial solutions will have one direction contracting and the others expanding, or vice-versa.

\subsubsection*{Axisymmetric solutions}

We now discuss the most general homogeneous and axisymmetric solution.
The most general anisotropic ansatz can be written as:
\begin{align}
\label{eq: ansatz}
&\phi^0 = f(t) + b_j x^j \; , \\
&\phi^i = A^i_{\; j} x^j + c^i(t) \; , \\
&\rmd s^2 = - \rmd t^2 + a_1^2(t) \rmd x^2 + a_2^2(t) \rmd y^2 + a_3^2(t) \rmd z^2 \; ,
\end{align}
where $b_j$ is a constant vector and $A^i_{\; j}$ is a constant matrix.

If we impose axial symmetry around the third direction, our ansatz should satisfy
\begin{align}
\label{eq: axisym}
&b_i = (0,0,q) \; , \qquad A^i_{\; j} = {\rm diag}(1,1,B)  \; , \qquad  c^i(t) = (0,0,c(t)) \\
& a_1(t) = a_2(t) = a(t) \; , \qquad a_3(t) = b(t) \; .
\end{align}
The matrix $\Sigma^\al_{\; \bt} = g^{\al \lam} \de_\lam \phi^a \de_\nu \phi^b \eta_{ab}$ is block diagonal and can be easily diagonalized.
Its eigenvalues are given by:
\begin{equation}
\sig_1^2 = \sig_2^2 = \frac{1}{a^2} \; , \qquad
\sig_3^2 = \frac{g + \sqrt{g^2 - h}}{b} \; , \qquad
\sig_4^2 = \frac{g - \sqrt{g^2 - h}}{b} \; ,
\end{equation}
where
\be
g \equiv \frac{1}{2} \left[ B^2 - q^2 + b^2 (\dot{f}^2 - \dot{c}^2) \right] \; , \qquad
h \equiv b^2 (q \dot{c} - B \dot{f})^2 \; .
\ee
We can derive two constraints by varying the action with respect to $f$ and $c$.
After some algebraic manipulations, we arrive at
\begin{align}
& \sqrt{\frac{B+q - \mu b (\dot{f}+ \dot{c})}{B - q - \mu b (\dot{f} - \dot{c})}} = \frac{a^2 (B+q) - \mu k_-}{b (3 a^2 - 2 a)} \, ,\\
& b^2 = \frac{[a^2 (B-q) - \mu k_+] [a^2 (B+q) - \mu k_-]}{(3 a^2-2a)^2} \; ,
\end{align}
where $\mu = {\rm sgn} (q \dot{c}-B\dot{f})$.

If we require that both $a$ and $b$ vanish when $t\to 0$, we must choose $k_+ = k_- =0$.
Now, from the second constraint, we find $B^2 > q^2$.
In order to have a positive $b$, the only possible choice would be $B+q < 0 \Rightarrow B < 0$.
We can prove that this is an inconsistent solution by looking at the square root.
First, we notice that the symmetry $\phi^a \to - \phi^a$ allows us to study only the case in which $B+q - \mu b (\dot{f}+\dot{c})>0$.
So, we would like to have
\be
\label{eq:sign}
\mu b (\dot{f}+\dot{c})<B+q<0 \; , \qquad \mu b (\dot{f}-\dot{c}) < B-q <0 \; .
\ee
Now, these conditions have no solution. In fact, they imply
\be
\mu \dot{f} <0 \; ,
\ee
which is absurd.
Indeed, we also see from \eqref{eq:sign} that $|\dot{f}| > |\dot{c}|$, which along with $|B|>|q|$ and $B<0$ implies that $\mu = {\rm sgn} \dot{f}$, so necessarily $\mu \dot{f} > 0$.
In conclusion, also in this case we cannot describe a Universe which expands starting from a small initial volume at early times.


\begin{thebibliography}{}

%\cite{deRham:2010kj}
\bibitem{drgt}
%\bibitem{dRGT}
 C.~de Rham, G.~Gabadadze and A.~J.~Tolley,
  ``Resummation of Massive Gravity,''
  Phys.\ Rev.\ Lett.\  {\bf 106}, 231101 (2011)
  [arXiv:1011.1232 [hep-th]].
  %%CITATION = PRLTA,106,231101;%%

%\cite{Boulware:1973my}
\bibitem{bd}
  D.~G.~Boulware and S.~Deser,
  ``Can gravitation have a finite range?,''
  Phys.\ Rev.\  D {\bf 6}, 3368 (1972).

 %\cite{deRham:2010ik}
%\bibitem{deRham:2010ik}
\bibitem{drg2}
C.~de Rham and G.~Gabadadze,
  ``Generalization of the Fierz-Pauli Action,''
  Phys.\ Rev.\  D {\bf 82} (2010) 044020
  [arXiv:1007.0443 [hep-th]].


%\cite{Hassan:2011hr}
\bibitem{hr}
  S.~F.~Hassan, R.~A.~Rosen,
  ``Resolving the Ghost Problem in non-Linear Massive Gravity,''
  [arXiv:1106.3344 [hep-th]].


\bibitem{drgt_res2}
C.~de Rham, G.~Gabadadze and A.~J.~Tolley,
  ``Ghost free Massive Gravity in the St\"uckelberg language,''
  arXiv:1107.3820 [hep-th].


\bibitem{helicity}
%\cite{deRham:2011qq}
%\bibitem{deRham:2011qq}
  C.~de Rham, G.~Gabadadze and A.~J.~Tolley,
  ``Helicity Decomposition of Ghost-free Massive Gravity,''
  arXiv:1108.4521 [hep-th].
  %%CITATION = ARXIV:1108.4521;%%

\bibitem{Arkady}
%\bibitem{Arkady}
%\cite{Vainshtein:1972sx}
% \bibitem{Vainshtein:1972sx}
  A.~I.~Vainshtein,
  ``To the problem of nonvanishing gravitation mass,''
  Phys.\ Lett.\  B {\bf 39}, 393 (1972).
  %%CITATION = PHLTA,B39,393;%%
%\bibitem{DDGV}
%\cite{Deffayet:2001uk}
%\bibitem{Deffayet:2001uk}

 \bibitem{DDGV}
%\cite{Deffayet:2001uk}
%\bibitem{Deffayet:2001uk}
  C.~Deffayet, G.~R.~Dvali, G.~Gabadadze and A.~I.~Vainshtein,
  ``Nonperturbative continuity in graviton mass versus perturbative
  discontinuity,''
  Phys.\ Rev.\  D {\bf 65}, 044026 (2002)
  [arXiv:hep-th/0106001].
  %%CITATION = PHRVA,D65,044026;%%

\bibitem{lpr}
  M.~A.~Luty, M.~Porrati and R.~Rattazzi,
  ``Strong interactions and stability in the DGP model,''
  JHEP {\bf 0309}, 029 (2003)
  [arXiv:hep-th/0303116].
  %%CITATION = JHEPA,0309,029;%%

\bibitem{nr}
  A.~Nicolis, R.~Rattazzi,
  ``Classical and quantum consistency of the DGP model,''
  JHEP 0406 (2004) 059
  [hep-th/0404159].
  %%CITATION = JHEPA,0406,059;%%


 %\cite{Nicolis:2008in}
\bibitem{nrt}
  A.~Nicolis, R.~Rattazzi and E.~Trincherini,
  ``The galileon as a local modification of gravity,''
  Phys.\ Rev.\  D {\bf 79}, 064036 (2009)
  [arXiv:0811.2197 [hep-th]].
  %%CITATION = PHRVA,D79,064036;%%


%\cite{Koyama:2011xz}
\bibitem{knt}
  K.~Koyama, G.~Niz, G.~Tasinato,
  ``Analytic solutions in non-linear massive gravity,''
[arXiv:1103.4708 [hep-th]],  [arXiv:1104.2143 [hep-th]].

 %\cite{Chkareuli:2011t
\bibitem{cp}
  G.~Chkareuli, D.~Pirtskhalava,
  ``Vainshtein Mechanism In $\Lambda_3$ - Theories,''
  [arXiv:1105.1783 [hep-th]].

 \bibitem{vdvz} H.~van Dam and M.~J.~G.~Veltman,
  ``Massive And Massless Yang-Mills And Gravitational Fields,''
  Nucl.\ Phys.\  B {\bf 22}, 397 (1970); V.~I.~Zakharov,
``Linearized gravitation theory and the graviton mass,''JETP Lett.\  {\bf 12}, 312 (1970)
 [Pisma Zh.\ Eksp.\ Teor.\ Fiz.\  {\bf 12}, 447 (1970)].
   %%CITATION = NUPHA,B22,397;%%


%\cite{deRham:2010tw}
%\bibitem{d}
\bibitem{drghp}
C.~de Rham, G.~Gabadadze, L.~Heisenberg and D.~Pirtskhalava,
  ``Cosmic Acceleration and the Helicity-0 Graviton,''
  Phys.\ Rev.\  D {\bf 83} (2011) 103516
  [arXiv:1010.1780 [hep-th]].

%\cite{covariant}
\bibitem{covariant}
  C.~de Rham, L.~Heisenberg,
  ``Cosmology of the Galileon from Massive Gravity,''
  Phys.\ Rev.\  {\bf D84}, 043503 (2011).
  [arXiv:1106.3312 [hep-th]].


\bibitem{theo}
%\cite{Nieuwenhuizen:2011sq}
%\bibitem{Nieuwenhuizen:2011sq}
  T.~.M.~Nieuwenhuizen,
  ``Exact Schwarzschild-de Sitter black holes in a family of massive gravity models,''
 [arXiv:1103.5912 [gr-qc]].


\bibitem{volkov}
%\cite{Chamseddine:2011bu}
%\bibitem{Chamseddine:2011bu}
  A.~H.~Chamseddine, M.~S.~Volkov,
  ``Cosmological solutions with massive gravitons,''
[arXiv:1107.5504 [hep-th]].

%\cite{ArkaniHamed:2002sp}
\bibitem{ags}
  N.~Arkani-Hamed, H.~Georgi, M.~D.~Schwartz,
  ``Effective field theory for massive gravitons and gravity in theory space,''
  Annals Phys.\  {\bf 305}, 96-118 (2003).
  [hep-th/0210184].

%\cite{Dubovsky:2004sg}
\bibitem{sergei}
  S.~L.~Dubovsky,
  ``Phases of massive gravity,''
  JHEP {\bf 0410}, 076 (2004).
  [hep-th/0409124].

%\cite{Afshordi:2006ad}
\bibitem{cuscuton}
  N.~Afshordi, D.~J.~H.~Chung, G.~Geshnizjani,
  ``Cuscuton: A Causal Field Theory with an Infinite Speed of Sound,''
  Phys.\ Rev.\  {\bf D75}, 083513 (2007).
  [hep-th/0609150].


\bibitem{degrav}
N.~Arkani-Hamed, S.~Dimopoulos, G.~Dvali and G.~Gabadadze,
  ``Nonlocal modification of gravity and the cosmological constant problem,''
  arXiv:hep-th/0209227; 
  %%CITATION = HEP-TH/0209227;%% 
G.~Dvali, S.~Hofmann and J.~Khoury,
  ``Degravitation of the cosmological constant and graviton width,''
  Phys.\ Rev.\  D {\bf 76}, 084006 (2007)
  [arXiv:hep-th/0703027].
  %%CITATION = PHRVA,D76,084006;%%


%\cite{Creminelli:2009mu}
\bibitem{guido}
  P.~Creminelli, G.~D'Amico, J.~Norena, L.~Senatore, F.~Vernizzi,
  ``Spherical collapse in quintessence models with zero speed of sound,''
  JCAP {\bf 1003}, 027 (2010).
  [arXiv:0911.2701 [astro-ph.CO]].



%\cite{Baumann:2010tm}
\bibitem{baumann}
  D.~Baumann, A.~Nicolis, L.~Senatore, M.~Zaldarriaga,
  ``Cosmological Non-Linearities as an Effective Fluid,''
[arXiv:1004.2488 [astro-ph.CO]].

%\cite{Fierz:1939ix}
\bibitem{fp}
  M.~Fierz and W.~Pauli,
  ``On relativistic wave equations for particles of arbitrary spin in an
  electromagnetic field,''
  Proc.\ Roy.\ Soc.\ Lond.\  A {\bf 173}, 211 (1939).











\end{thebibliography}
\end{document}